# Controlled synthesis of Mo$_x$W$_{1-x}$Te$_2$ atomic layers with emergent quantum states

Ya Deng[†,Δ], Peiling Li[‡,§,Δ], Chao Zhu[†,Δ], Jiadong Zhou[†], Xiaowei Wang[†], Jian Cui[‡], Xue Yang[‡,§], Li Tao[∥], Qingsheng Zeng[†], Ruihuan Duan[†], Qundong Fu[†], Chao Zhu[†], Jianbin Xu[∥], Fanming Qu[‡,§,#], Changli Yang[‡,§], Xiunian Jing[‡,#], Li Lu[‡,§,#], Guangtong Liu[‡,§,#*] and Zheng Liu[†*]

[†]*Center for Programmable Materials, School of Materials Science & Engineering, Nanyang Technological University, Singapore 639798, Singapore*

[‡]*Beijing National Laboratory for Condensed Matter Physics, Institute of Physics, Chinese Academy of Sciences, Beijing 100190, China*

[§]*University of Chinese Academy of Sciences, Beijing 100049, China*

[∥]*Department of Electronic Engineering, The Chinese University of Hong Kong, Shatin, N.T., Hong Kong SAR, China*

[#]*Songshan Lake Materials Laboratory, Dongguan, Guangdong 523808, China*

ABSTRACT: Recently, new states of matter like superconducting or topological quantum states were found in transition metal dichalcogenides (TMDs) and manifested themselves in a series of exotic physical behaviors. Such phenomena have been demonstrated to exist in a series of transition metal tellurides including MoTe$_2$, WTe$_2$ and alloyed Mo$_x$W$_{1-x}$Te$_2$. However, the behaviors in the alloy system have been rarely addressed due to their difficulty in obtaining atomic layers with controlled composition, albeit the alloy offers a great platform to tune the quantum states. Here, we report a facile CVD method to synthesize the Mo$_x$W$_{1-x}$Te$_2$ with controllable thickness and chemical composition ratios. The atomic structure of monolayer Mo$_x$W$_{1-x}$Te$_2$ alloy was experimentally confirmed by scanning transmission electron microscopy (STEM). Importantly, two different transport behaviors including superconducting and Weyl semimetal (WSM) states were observed in Mo-rich Mo$_{0.8}$W$_{0.2}$Te$_2$ and W-rich Mo$_{0.2}$W$_{0.8}$Te$_2$ samples respectively. Our results show that the electrical properties of Mo$_x$W$_{1-x}$Te$_2$ can be tuned by controlling the chemical composition, demonstrating our controllable CVD growth method is an efficient strategy to manipulate the physical properties of TMDCs. Meanwhile, it provides a perspective on further comprehension and shed light on the design of device with topological multicomponent TMDCs materials.

INTRODUCTION: Topological sates[1-3] have been observed in two-dimensional (2D) transition metal dichalcogenides (TMDCs),[4] making them a fertile ground for discovering new quasiparticles in condensed matter physics, such as Weyl and Dirac fermions.[5,6] For example, type-II Weyl semimetals (WSM) states have been demonstrated to exist in Mo$_x$W$_{1-x}$Te$_2$ system[7-9] due to their inherent semimetallic distorted octahedral phase with broken inversion symmetry.[10,11] Different from the standard (type-I) WSMs with point-like Fermi surfaces, the existence of tilted Weyl cones and Fermi arcs[5,12,13] has led to many exotic transport behaviors such as anisotropic magnetoconductance enhancement and extra quantum oscillations. Moreover, the superconducting and topological insulating state have recently been reported to exist in $MoTe_2$[14] and $WTe_2$,[15,16] respectively. If we can incorporate the superconductivity and topological insulating state in the alloyed $Mo_xW_{1-x}Te_2$ system, the long-sought topological superconductivity will be realized in a single material. The angle-resolved photoemission spectroscopy (ARPES) has shown that the Fermi arcs can be tuned by varying the Mo/W ratio[7] in $Mo_xW_{1-x}Te_2$ system. Therefore, the alloyed $Mo_xW_{1-x}Te_2$ system offers a great platform to tune the quantum states. However, previous work mainly focused on the phase structures of this $Mo_xW_{1-x}Te_2$ alloy are widely studied by Raman, XRD and *etc*.[17,18] It has reached a consensus that a 2H or 1T′ phase is dominated in Mo-rich sample, while $1T_d$ phase dominated in W-rich sample.[17-19] Besides, these research projects are mostly based on the $Mo_xW_{1-x}Te_2$ alloy synthesized by chemical vapor transport (CVT) method, which are generally bulk crystals, challenging to achieve controllable thickness and mass production. To further broaden the research potential of $Mo_xW_{1-x}Te_2$ alloy, the chemical vapor deposition (CVD) method which widely regarded as a facile technology in controllable synthesis and large-scale applications,[20] appears to be a promising alternative for fast and scalable production of thickness-controlled $Mo_xW_{1-x}Te_2$ alloy. Recently, Chen *et al.* reported the CVD growth of $Mo_xW_{1-x}Te_2$ film with the assist of hBN covered substrate, but probably due to the polycrystalline structure or substrate limit, no superconductivity or Weyl semimetal related properties were observed.[21] Therefore, obtaining high-quality atomic layers with controlled compositions of $Mo_xW_{1-x}Te_2$ alloy on normal silicon substrate remains challenging at present.

In this work, we report a CVD method to synthesize $Mo_xW_{1-x}Te_2$ alloy with different thicknesses and tunable stoichiometric ratio of cations Mo/W. The atomic structure of monolayer $Mo_xW_{1-x}Te_2$ alloy was experimentally confirmed by scanning transmission electron microscopy (STEM).

Based on the rapid growth strategy, the Mo$_x$W$_{1-x}$Te$_2$ alloy in a distorted octahedral phase can be obtained. The electronic properties of Mo$_x$W$_{1-x}$Te$_2$ with different stoichiometric ratios have also been investigated by the low-temperature transport measurements. As a preliminary character of type-II Weyl semimetal, the expected negative magnetoresistance was experimentally observed in Mo$_{0.2}$W$_{0.8}$Te$_2$ flake. Meanwhile, the superconducting transition behavior at 1.8 K was observed in Mo$_{0.8}$W$_{0.2}$Te$_2$ devices. In short, by tuning the Mo concentration, the superconductivity and chiral-anomaly-induced negative magnetoresistance can be realized in CVD-synthesized Mo$_x$W$_{1-x}$Te$_2$ alloy, revealing the high quality and great potentials of Mo$_x$W$_{1-x}$Te$_2$ as a platform for exploring new physical phenomena.

RESULTS AND DISCUSSION

The molten-salt assisted CVD method[20] has been used to synthesize the Mo$_x$W$_{1-x}$Te$_2$ alloys, with corresponding schematic diagrams of thermal CVD processes is shown in Figure S1 of the Supporting Information. We fixed the reaction time and gas flow as 2 min and 100 sccm Ar/ 5 sccm H$_2$, first explore the effect of reaction temperature. By tuning the reaction temperature from 700 °C to 850 °C, the obtained Mo$_x$W$_{1-x}$Te$_2$ atomic layers shown different morphologies, and more importantly, they were further confirmed with different composition ratio of Mo and W elements in Mo$_x$W$_{1-x}$Te$_2$. Then, by further optimizing the growth parameter, the Mo$_x$W$_{1-x}$Te$_2$ alloys with controllable morphologies can be obtained as shown in Figure 1 and Figure S3. More information about the sample growth is detailed in the "Materials and Method" section and Figures S1-S3. The morphologies of as-synthesized Mo$_x$W$_{1-x}$Te$_2$ alloys were characterized by optical microscopy and atomic force microscopy (AFM). Figure 1a is the schematic of Mo$_x$W$_{1-x}$Te$_2$ crystal structure, which illustrates the dopant W atoms (orange) in MoTe$_2$ lattice. Figures 1b and 1c show the optical images of mono- and bi-layer Mo$_x$W$_{1-x}$Te$_2$ with a length of ~100 μm and a width of ~10 μm. The ribbon-like morphologies suggest that the $Mo_xW_{1-x}Te_2$ is single crystallized in 1T′ or $1T_d$ phase, which is supported by the following Raman spectra and STEM characterization. Figures 1d and 1e exhibit a uniform polycrystalline monolayer film with a lateral size larger than 100 μm and a flake with different thicknesses. The sample with different morphology could simultaneously meet the requirements of different investigation such as large- area or thickness-dependent research. The thickness of $Mo_xW_{1-x}Te_2$ film is verified by AFM. The typical AFM height topography shown in Figure 1f indicates the measured thickness of $Mo_xW_{1-x}Te_2$ is of ~1.1 nm, although it's a bit larger than the actual single-layer thickness due to the small oxidized particles formed on the materials surface, it is completely consistent with the previous AFM measured thickness of monolayer $WTe_2$ and $MoTe_2$.[22,23] Therefore, the measured thickness of 1.1 nm can be regarded as monolayer structure. Meanwhile, the smooth and uniform surface morphology reveals a good crystal quality.

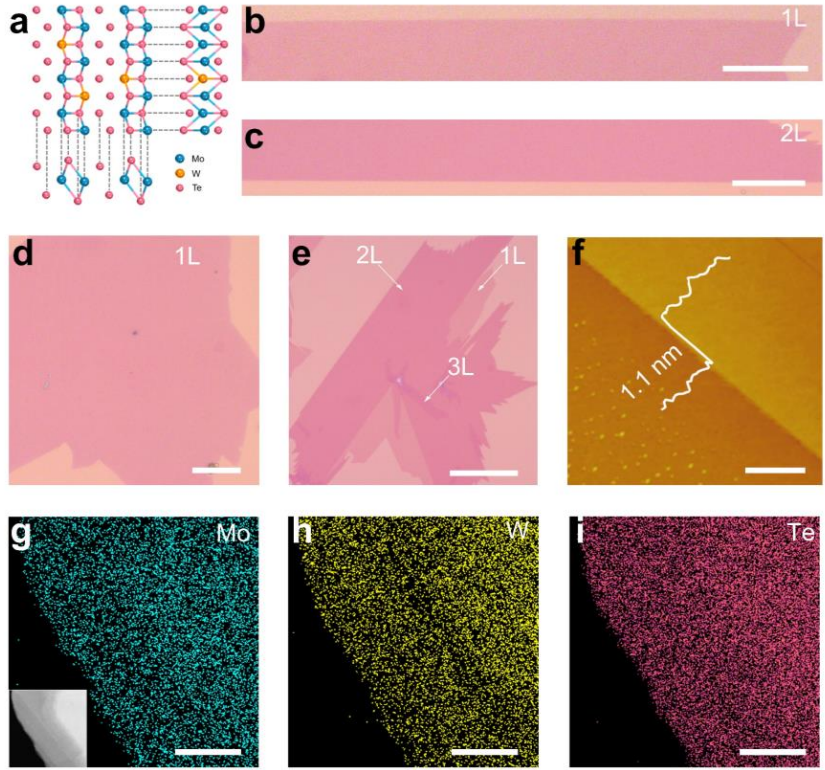

**Figure 1.** Morphologies and characterizations of 1T′ phase $Mo_xW_{1-x}Te_2$ atomic layers. (a) Structural model of 1T′ $Mo_xW_{1-x}Te_2$ viewed from front, side, and top directions. (b, c) Optical images of typical single crystalline monolayer (b) and bilayer $Mo_xW_{1-x}Te_2$ (c) with the length around 100 μm deposited on the $SiO_2$/Si substrates. (d, e) Optical images of a large uniform monolayer $Mo_xW_{1-x}Te_2$ film (d) and flake (e) containing different layer numbers. (f) A representative atomic force microscopy (AFM) image at the flake edge shows the typical monolayer thickness is 1.1 nm. (g-i) Typical EDX elemental mappings of Mo (blue), W(yellow) and Te (red) in a $Mo_{0.5}W_{0.5}Te_2$ flake (morphology shown in the SEM image inserted in (g)). Scale bars: 10 μm in (b, c), 20 μm in (d), 50 μm in (e), 1 μm in (f) and 50 nm in (g-i).

Then the energy-dispersive X-ray spectroscopy (EDX) was employed to analyze the compositions of as-synthesized $Mo_xW_{1-x}Te_2$ alloy. The EDX element distributions shown in Figures 1g-i and corresponding spectrum in Figure S4 confirm the synthesized material is composed of uniformly distributed Mo, W and Te atoms. Afterwards, we synthesized the $Mo_xW_{1-x}Te_2$ under different growth temperatures, as described in "Materials and Method" section. The growth temperature was found to be a key factor to determine the atomic ratio of Mo and W in the alloy. In previous works, limited by the bulk crystal structure, most of them used EDX to characterize the stoichiometry of the alloy materials.[17,18,24] However, in principle, when compared to the qualitative element characterization technique like EDX, XPS can provide more accurate quantitative characterization results. The CVD-grown $Mo_xW_{1-x}Te_2$ thin film with a large size is more suitable for standard XPS analysis of the element ratio. The X-ray photoelectron spectroscopy (XPS) was employed to analyze the compositions of the alloy. The representative XPS survey spectra shown in Figure S5 indicate the observed peak positions are consistent with the binding energies of Mo, W and Te element,[22,25] confirming the existence of each constituent element in the $Mo_xW_{1-x}Te_2$ alloy. Furthermore, as the peak intensity in XPS spectra reflects the relative amount of constituent element, the value of $x$ in $Mo_xW_{1-x}Te_2$ can be estimated by comparing the intensities of XPS signal. The XPS spectra from $Mo_xW_{1-x}Te_2$ alloys with different growth temperatures are shown in Figures 2a and S6. Four peaks located at 31 eV, 34 eV, 228 eV and 231 eV can be clearly resolved, corresponding to W 4f 7/2, W 4f 5/2, Mo 3d 5/2 and Mo 3d 3/2, respectively.[22,25] The peaks of Te 3d are shown in Figure S6, the component ratios of Te in the alloy are all calculated around 2, indicating a stoichiometric $Mo_xW_{1-x}Te_2$. As the growth temperature increases, the signal from Mo becomes weak, while the signal from W is enhanced. The corresponding $x$ are calculated to be 0.85, 0.74, 0.52, 0.17 respectively. This result is summarized in Figure 2b, where the Mo concentration $x$ and growth temperature are highly correlated: low growth temperature tends to form the alloy with high Mo atomic concentration.

Besides, the XPS spectra from $Mo_xW_{1-x}Te_2$ alloys with different thicknesses are shown in Figures S7, suggesting good uniformity of synthesized $Mo_xW_{1-x}Te_2$ alloy.

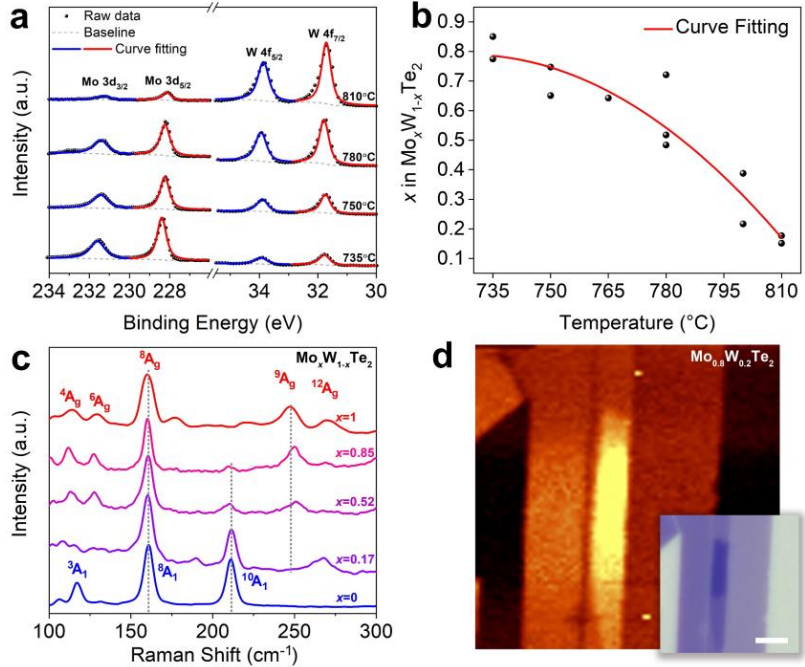

**Figure 2**. XPS and Raman characterization of the $Mo_xW_{1-x}Te_2$ alloy with different Mo concentrations. (a) XPS spectra of the as-grown $Mo_xW_{1-x}Te_2$ alloy at different growth temperatures. Stoichiometric ratio of cations Mo and W are calculated based on the intensity ratio of strongest peaks (fitted as red line), the Mo concentration $x$ of 0.85, 0.74, 0.52, 0.17 in $Mo_xW_{1-x}Te_2$ corresponds to the growth temperature of 735°C, 750°C, 780°C, 810°C, respectively. (b) The relationship between the Mo concentration $x$ and growth temperature of as grown $Mo_xW_{1-x}Te_2$ alloy. The Mo concentration $x$ is found to be temperature dependent. (c) Raman spectra of $Mo_xW_{1-x}Te_2$ trilayers with different $x$. The values of $x$ are marked on the right hand of each spectrum ($x = 0$ is pure $MoTe_2$ and $x=1$ is pure $WTe_2$). The dashed lines are guides for the eye. (d) Raman intensity mapping ($^9A_g$) of $Mo_{0.8}W_{0.2}Te_2$ flake from the region indicated by the inserted optical image, confirming the uniform distribution of Mo. Scale bar: 3 μm in the inset of (d).

Relying on Raman characterization technology, we further discovered that the variation of stoichiometric ratio in the alloy is also reflected in Raman spectra, as shown in Figure 2c. The trilayer $Mo_xW_{1-x}Te_2$ were used as the objects of Raman characterization, as the trilayers are more stable while their Raman features are also enough to represent thin layer samples. The marked value of $x$ on the right hand of each spectrum denotes the stoichiometric ratio of Mo element in $Mo_xW_{1-x}Te_2$ alloy. The characteristic peaks shown in the spectra are marked with corresponding vibration modes. The $^{10}A_1$ (214 cm$^{-1}$) mode and $^{12}A_g$ (268 cm$^{-1}$) modes are denoted as the fingerprint vibration modes in 1T′ $WTe_2$ and $MoTe_2$ respectively.[20,26-27] Therefore, the simultaneous existence of vibration modes of $^{10}A_1$ and $^{12}A_g$ in the spectra confirms the coexistence of Mo and W in the alloy once again. This phenomenon is inconsistent with previous reported Raman spectra of bulk $Mo_xW_{1-x}Te_2$ alloy where two characteristic peaks were not observed at the same time except a sudden change in a certain ratio.[17] It may be attributed to the layer-dependent behaviors of peaks, such as the $A^5_1$ peak located at 130 cm$^{-1}$ in $WTe_2$, which was obvious in bulk crystal but became weak or almost disappeared in thin layer.[22,23,28] This is also the reason why it is not feasible to use this peak to characterize the doping state in our alloy samples. In the spectrum of $Mo_{0.52}W_{0.48}Te_2$, the $^{10}A_1$ and $^9A_g$ peaks broaden significantly, possibly being attributed to heavily doped atomic structure. The intensity changes of $^{10}A_1$ and $^9A_g$ peaks are found to be well matched with the marked $x$ obtained from previous XPS results in Figure 2a. In addition, the variation of $x$ also embodies in the peak position of $^9A_g$ and $^{12}A_g$ (Figure S8a and S8b). Further Raman intensity mapping of $^{10}A_1$ and $^9A_g$ modes in $Mo_{0.8}W_{0.2}Te_2$ alloy reveals homogeneous nature across the region with the same layer and indicates a uniform distribution of W and Mo in the $Mo_xW_{1-x}Te_2$ (Figure 2d and Figure S8c). For the $Mo_xW_{1-x}Te_2$ with different layers, the peak position of $^{12}A_g$ modes differs significantly, as shown in Figure S8d. This observation agrees well with the intrinsic Raman feature of MoTe$_2$.[26]

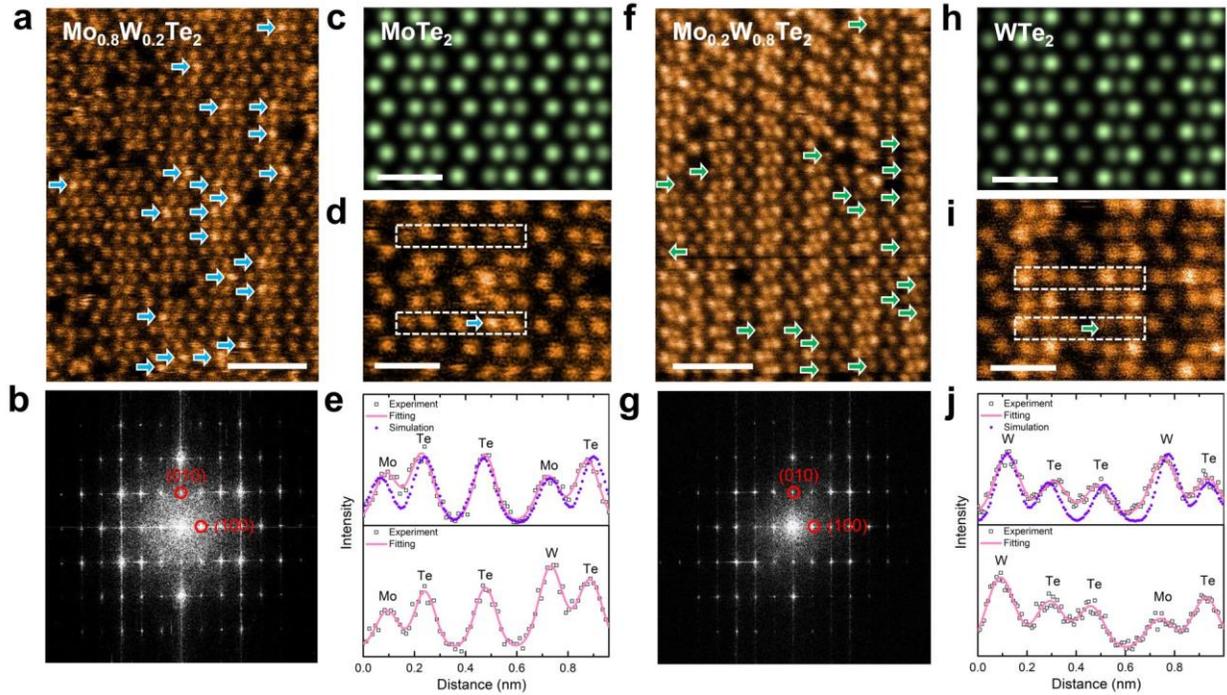

**Figure 3**. Atomic resolution STEM characterization of Mo$_x$W$_{1-x}$Te$_2$ alloy monolayers. (a) ADF-STEM image of monolayer Mo$_{0.8}$W$_{0.2}$Te$_2$ with the doped Mo atoms highlighted by arrows. (b) FFT of monolayer Mo$_{0.8}$W$_{0.2}$Te$_2$ STEM image. (c) Simulated ADF-STEM image of monolayer MoTe$_2$. (d) Enlarged ADF-STEM image of monolayer Mo$_{0.8}$W$_{0.2}$Te$_2$. (e) Line intensity files of the un-doped atom chains in both (c) and (d) (upper dashed rectangle), and doped atom chain in d (lower dashed rectangle) highlighted by rectangles. (f) ADF-STEM image of monolayer Mo$_{0.2}$W$_{0.8}$Te$_2$ with the doped tungsten atoms highlighted by arrows. (g) FFT of monolayer Mo$_{0.2}$W$_{0.8}$Te$_2$ STEM image. (h) Simulated ADF-STEM image of monolayer WTe$_2$. (i) Enlarged ADF-STEM image of monolayer Mo$_{0.2}$W$_{0.8}$Te$_2$. (j) Line intensity files of the un-doped atom chains in both (h) and (i) (upper dashed rectangle), and doped atom chain in (i) (lower dashed rectangle). Scale bars: 1nm in (a, f) and 0.5 nm in (c, d, h, i).

In order to clarify the atomic structure of $Mo_xW_{1-x}Te_2$, STEM characterization was performed. Different from the STEM of multi-layer samples,[29] the judgement (direct observations) of doping state in the alloy based single-layer sample is considered to be more accurate, because the influence of atomic overlap can be intentionally excluded. In this work, we examined the monolayer Mo-rich ($Mo_{0.8}W_{0.2}Te_2$) and W-rich ($Mo_{0.2}W_{0.8}Te_2$) alloys using STEM, as shown in Figures 3a and 3f. The fast Fourier transformation (FFT) patterns of large-area annular dark-field (ADF) STEM images confirm the rectangular structure of $Mo_xW_{1-x}Te_2$ unit cell, with (100) d-spacing of 0.63 nm for $Mo_{0.8}W_{0.2}Te_2$ and 0.64 nm for $Mo_{0.2}W_{0.8}Te_2$ respectively, as presented in Figures 3b and 3g. We found that, as indicated by the arrows in Figures 3a and 3f, the tungsten and molybdenum dopant are evenly distributed among each alloy and the doping concentration is always around 20% ($x \sim 0.2$) at different locations (more images are provided in Figure S9), indicating the high quality and uniformity of our samples. Meanwhile, we find the doping concentration characterized by the STEM agrees well with the XPS results. The observed holes and broken areas in Figures 3a and 3f are damaged lattice during STEM characterization because these monolayer alloys are vulnerable to electron irradiation (Figure S10). The enlarged ADF-STEM image of monolayer $Mo_{0.8}W_{0.2}Te_2$ in Figure 3d resembles that of $MoTe_2$ in the simulated image of Figure 3c except for some dopant tungsten atoms. Figure 3e shows the line intensity file along *a* axis of simulated $MoTe_2$, the undoped region of $Mo_{0.8}W_{0.2}Te_2$ (upper dashed rectangle in Figure 3d) and the doped region of $Mo_{0.8}W_{0.2}Te_2$ (lower dashed rectangle in Figure 3d). It suggests that the intensity distribution of Mo-Te-Te-Mo-Te atomic chain in the un-doped region is consistent with the simulation, meanwhile the doped tungsten atoms with higher intensity can be clearly distinguished in the doped Mo-Te-Te-W-Te atomic chain. Similarly, doped molybdenum atoms can be recognized in $Mo_{0.2}W_{0.8}Te_2$ monolayer (Figures 3h-j). Besides, it should be noticed that the 1T' and 1T$_d$ are both distorted octahedral phase which share same monolayer structures but different stacking along *c*-axis (Figures S11). Due to the similarity between 1T' and 1T$_d$, although many efforts have been devoted to distinguish these two phases through research on Raman, XRD, cross-section TEM and ARPES, *etc.*,[17,18,24,30] distinguishing them from each other becomes very difficult. But according to the intrinsic phase structure of MoTe$_2$ and WTe$_2$,[22,31-33] and previous research in Mo$_x$W$_{1-x}$Te$_2$,[17-19] it has reached a consensus that a 1T' phase is dominated in Mo-rich sample and 1T$_d$ phase dominated in W-rich sample, while a phase mixing behavior is presented in most composition ratios of Mo$_x$W$_{1-x}$Te$_2$ alloys. Therefore, the Mo-rich Mo$_{0.8}$W$_{0.2}$Te$_2$ and W-rich Mo$_{0.2}$W$_{0.8}$Te$_2$ samples we studied can be reasonably regarded as 1T' and 1T$_d$ phase respectively.

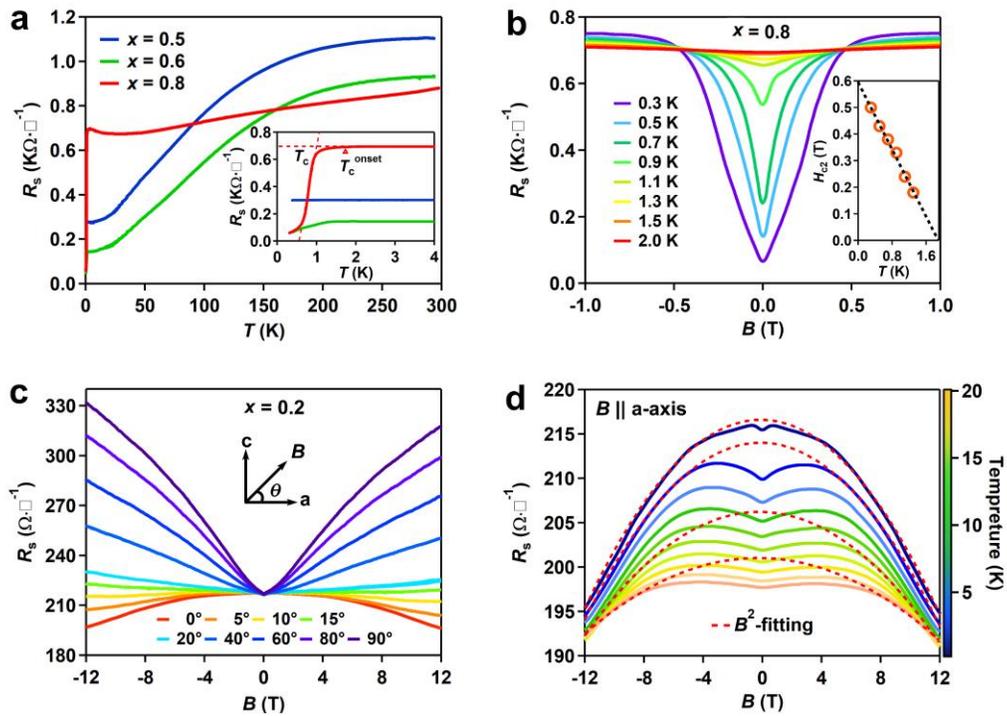

**Figure 4**. Transport characterization of the as-synthesized Mo$_x$W$_{1-x}$Te$_2$ sample. (a) Temperature dependence of longitudinal sheet resistance $R_s$ for Mo$_x$W$_{1-x}$Te$_2$ flakes with *x*=0.5, 0.6 and 0.8. The inset of (a) plots a zoom of the plot at low temperature. The triangle and dashed lines give $T_c^{onset}$=1.8 K and $T_c$=1.0 K. (b) Magnetic field dependence of $R_s$ for Mo$_{0.8}$W$_{0.2}$Te$_2$ sample

measured at selected temperatures. The inset shows the upper critical field $H_{c2}(T)$ for $Mo_{0.8}W_{0.2}Te_2$, which can be well fitted by a linear relation (black dashed line). (c) Magnetic field dependence of longitudinal sheet resistance $R_s$ for W-rich $Mo_{0.2}W_{0.8}Te_2$ film at $T$=0.3 K collected at different tilt angles $\theta$. The inset is a schematic of the tilt experiment setup, where $a$ and $c$ represent the crystallographic axes, $\theta$ is the tilt angle between the magnetic field $B$ and the positive direction of $a$-axis. (d) Magnetic field dependence of in-plane sheet resistance $R_s(B)$ for $Mo_{0.2}W_{0.8}Te_2$ at different temperatures in parallel magnetic field $\theta$=0°. The red dashed lines are the fitting curves with $R_s(B) \propto -B^2$ for four typical temperatures.

To further characterize the quality of as-synthesized $Mo_xW_{1-x}Te_2$ crystals of different compositions, we carried out low-temperature transport experiments. Interestingly, we observed two different transport behaviors in Mo-rich ($x$>0.5) and W-rich ($x$≤0.5) samples, indicating our controllable CVD growth method is an efficient strategy to manipulate the physical properties. Figure 4a shows the temperature dependence of longitudinal sheet resistance $R_s(T)$ for $Mo_xW_{1-x}Te_2$ flakes with $x$=0.5, 0.6 and 0.8 at zero-magnetic field. Clearly, all the three samples exhibit metallic conducting behavior at high temperatures ($T$>4 K). As shown in the inset of Figure 4a, we find that $Mo_xW_{1-x}Te_2$ flakes with $x$ =0.5 reaches a residual resistance below $T$=4.0 K due to defects or impurity, while the Mo-rich $Mo_xW_{1-x}Te_2$ samples ($x$=0.6 and 0.8) show a broad superconducting transition. Furthermore, the superconducting fraction $[R(T_c^{onset}) - R(0.3\ K)]/R(T_c^{onset})$ is found to be strongly dependent on the Mo concentration, which increases from 56% to 91% with $x$ increased from 0.6 to 0.8. Here, $T_c^{onset}$ is defined as the temperature where $R_s$ begins to deviate from the normal state resistance, as shown in the inset of Figure 4a. To further explore the dimensionality of superconductivity in $Mo_{0.8}W_{0.2}Te_2$, we measured the isothermal sheet resistance as a function of the magnetic field $B$ at selected temperatures, as shown in Figure 4b. Here, the upper critical field $H_{c2}$ is defined as the field where $R_s$ begins to drop (the onset of superconductivity). With increasing temperature, the superconducting transition shifts to lower temperatures and systematically broadens. Finally, the superconductivity is completely suppressed at $T$ = 2.0 K. The inset of Figure 4b shows the extracted upper critical field $H_{c2}$ as a function of $T$, which can be well fitted by a linear relation. This is a characteristic of 2D superconductors[34-36] and can be explained by the Ginzburg-Landau (GL) theory,[37] $H_{c2}(T) = \frac{\phi_0}{2\pi \xi_{GL}(0)^2}(1 - \frac{T}{T_c})$, where $\xi_{GL}(0)$ is the zero-temperature GL coherence length and $\phi_0$ is the magnetic flux quantum. The best fitting yields $\xi_{GL}(0)$ =23 nm, which is larger than the sample thickness $d$=13.8 nm, demonstrating the measured $Mo_{0.8}W_{0.2}Te_2$ flake is a 2D superconductor (See Note SI for more devices). We note that similar broad superconducting transition has been reported in two-dimensional (2D) superconductors, such as thin film YBCO,[38] few-layer $NbSe_2$,[34] few-layer TiO,[39] gated-ZrNCl,[40] -$MoS_2$[41] and -$SnSe_2$.[42] The possible intrinsic mechanism includes thermal fluctuation, thermally assisted flux flow, thermal creep or quantum metallic states,[34,38-42] which is an interesting question and needs to be studied in future.

In stark contrast to $Mo_{0.8}W_{0.2}Te_2$, we observed unsaturated magnetoresistance ( MR = $\frac{R_S(B)-R_S(0)}{R_S(0)} \times 100\%$) of ~57% in $Mo_{0.2}W_{0.8}Te_2$ at $T$=0.3 K in a perpendicular magnetic field of 12 T. Figure 4c shows the results of *in situ* tilt experiments performed at $T$=0.3 K. Interestingly, we find that a negative MR occurs within $\theta$ < 15° and reaches a maximum at $\theta$ = 0°. For a type-II Weyl semimetal, recent studies[33,43-46] have shown that chiral-anomaly-induced negative MR should be observed when the magnetic field is rotated to be parallel to the applied current and the tilting direction of Weyl cones. Therefore, the direction-dependent negative MR observed in $Mo_{0.2}W_{0.8}Te_2$ is a strong signature of a type-II Weyl semimetal. To clarify the origin of the negative MR, we performed quantitative analysis of the low-temperature in-plane MR ($\theta$=0°), as shown in

Figure 4d. From 0.3 K to 20 K, the in-plane MR are all negative. The low-field cusp-like dip is regarded to arise from weak antilocalization (WAL)[47,48] (see details in Figure S14 and Note SII). Moreover, the high-field data can be described by $R_s(B) \propto -B^2$, which is consistent with the theoretical expectation for chiral-anomaly-induced negative MR,[33,43-46] and support $Mo_{0.2}W_{0.8}Te_2$ is a Weyl semimetal. In addition, by tuning the sample thickness of $Mo_{0.2}W_{0.8}Te_2$, the coexistence of electron-electron interaction and weak antilocalization is observed in thinner samples (Figure S15 and Note SII). The observations of superconductivity and chiral-anomaly-induced negative resistance in Mo- and W-rich samples prove that our samples are high quality.

CONCLUSIONS

In conclusion, we have developed the salt-assistant CVD method to directly synthesize high-quality $Mo_xW_{1-x}Te_2$ alloy. The composition ratio of Mo and W in synthesized $Mo_xW_{1-x}Te_2$ atomic layers can be controlled by customizing growth temperatures. The AFM, XPS, Raman and STEM characterizations have revealed good quality and high uniformity in the alloy, as evidenced by observing superconductivity and Weyl semimetal state in Mo- and Wo-rich samples, respectively. Moreover, the critical Mo concentration where the alloy exhibit superconducting behavior is found to be above $x$=0.5. Our results show that a facile CVD method is capable of synthesizing the $Mo_xW_{1-x}Te_2$ alloy with controllable thickness and composition ratio, which provides opportunities to investigate exotic physical phenomenon, including topological superconductivity by incorporating superconductivity and topology in $Mo_xW_{1-x}Te_2$ system.

MATERIALS AND METHODS

**Synthesis of atomically thin $Mo_xW_{1-x}Te_2$ alloy:** We use the powders of $WO_3$, $MoO_3$ (Sigma) as the precursor for growing the $Mo_xW_{1-x}Te_2$ alloys by ambient pressure CVD method (Figure S1).

In addition, the grains of NaCl were added to decrease the melting point of precursors as reported by previous work of our group.[20] As the product is sensitive to hydrogen, excessive of which may cause a sharp decrease of Mo in the synthesized alloy, we used the mixed gas of $H_2$/Ar with 5 and 100 sccm as the carrier gas. The growth temperature varied from 700 °C to 850 °C and held for 2-10 mins. The heating rate is 50 °C/min for all processes. Then the $Mo_xW_{1-x}Te_2$ atomic layers were obtained straightforward with different morphologies (Figure S2). Moreover, by further optimizing the growth parameters under the same temperature, the morphologies can be controlled as shown in Figure S3. The $MoTe_2$ and $WTe_2$ samples used for comparative Raman experiments were synthesized by a salt-assist method reported in ref 22.

**AFM, XPS and Raman Characterization:** AFM analysis was performed using the Asylum Research Cypher AFM in tapping mode. The XPS spectra were measured by the Kratos AXIS Supra spectrometer with a dual anode Al-Kα (1486.6 eV) X-ray monochromatic source. The collected XPS spectra were calibrated with $C_{1s}$ binding energy of 284.8 eV. Raman measurements were performed by WITEC alpha 200R Confocal Raman system with 532 nm excitation laser. To avoid samples overheating, the adopted laser powers were kept below 1 mW.

**STEM Characterization:** The STEM samples were prepared by wet-etching method. The poly methyl methacrylate (PMMA) was spin-coated on the SiO2/Si substrate deposited with $Mo_xW_{1-x}Te_2$ flakes. The obtained substrate was then immersed in the NaOH solution (1 M) and DI water sequentially to etch the SiO2 layers and remove the KOH residue. Afterwards, a film was found floating on the water and was fished by a TEM grid (Quantifoil Au grid) to get a PMMA/ $Mo_xW_{1-x}Te_2$/Grid structure. The mentioned structure was dropped into acetone and then baked at 250 °C in vacuum for 60 mins to remove the PMMA.

**Device fabrication and transport measurement:** High-quality $Mo_xW_{1-x}Te_2$ flakes were directly deposited on $SiO_2$/Si substrate. The Hall-bars devices with Ti/Au (5/70 nm) electrodes were fabricated on synthesized $Mo_xW_{1-x}Te_2$ flakes using standard e-beam lithography (EBL) process followed by electron-beam evaporation and lift-off process. Low-temperature transport experiments were carried out with a standard four-terminal method from room temperature to 0.3 K in a top-loading Helium-3 refrigerator with a 15 T superconducting magnet.

ASSOCIATED CONTENT

**Supporting Information**.

The Supporting Information is available free of charge on the ACS Publications website. Scheme of CVD process, optical images of synthesized materials, supplementary EDX, XPS and Raman spectra, more STEM images of monolayers, transport characterization of $Mo_xW_{1-x}Te_2$ devices (PDF).

AUTHOR INFORMATION

**Corresponding Author**

*z.liu@ntu.edu.sg

*gtliu@iphy.ac.cn

**Author Contributions**

The manuscript was written through contributions of all authors. All authors have given approval to the final version of the manuscript. $^{\triangle}$These authors contributed equally.

**Notes**

The authors declare no competing financial interest.

ACKNOWLEDGMENT

Z.L. acknowledges supports from National Research Foundation Singapore programme (NRF-CRP22-2019-0007 and NRF-CRP21-2018-0007) and Singapore Ministry of Education AcRF Tier 3 Programme 'Geometrical Quantum Materials' (MOE2018-T3-1-002). T.G.L. acknowledges supports from National Basic Research Program of China from the MOST (2016YFA0300601), National Natural Science Foundation of China (11527806, 11874406 and 92065203), Beijing Municipal Science & Technology Commission of China (Z191100007219008), Beijing Academy of Quantum Information Sciences (Y18G08), Strategic Priority Research Program of the Chinese Academy of Sciences (XDB33010300), the Synergic Extreme Condition User Facility.

REFERENCES

(1) Chhowalla, M.; Shin, H. S.; Eda, G.; Li, L. J.; Loh, K. P.; Zhang, H. The Chemistry of Two-Dimensional Layered Transition Metal Dichalcogenide Nanosheets. *Nat. Chem.* **2013,** *5*, 263.

(2) Duan, X.; Wang, C.; Pan, A.; Yu, R.; Duan, X. Two-Dimensional Transition Metal Dichalcogenides as Atomically Thin Semiconductors: Opportunities and Challenges. *Chem. Soc. Rev.* **2015,** *44*, 8859-8876.

(3) Choi, W.; Choudhary, N.; Han, G. H.; Park, J.; Akinwande, D.; Lee, Y. H. Recent Development of Two-Dimensional Transition Metal Dichalcogenides and Their Applications. *Mater. Today* **2017,** *20*, 116-130.

(4) Manzeli, S.; Ovchinnikov, D.; Pasquier, D.; Yazyev, O. V.; Kis, A. 2D Transition Metal Dichalcogenides. *Nat. Rev. Mater.* **2017,** *2*, 17033.

(5) Wan, X.; Turner, A. M.; Vishwanath, A.; Savrasov, S. Y. Topological Semimetal and Fermi-Arc Surface States in the Electronic Structure of Pyrochlore Iridates. *Phys. Rev. B* **2011,** *83*, 205101.

(6) Xu, S. Y.; Belopolski, I.; Alidoust, N.; Neupane, M.; Bian, G.; Zhang, C.; Sankar, R.; Chang, G.; Yuan, Z.; Lee, C.C.; Huang, S. M.; Zheng, H.; Ma, J.; Sanchez, D. S.; Wang, B.; Bansil, A.; Chou, F.; Shibayev, P. P.; Lin, H.; Jia, S.; *et al*. Discovery of a Weyl Fermion Semimetal and Topological Fermi Arcs. *Science* **2015,** *349*, 613-617.

(7) Chang, T. R.; Xu, S. Y.; Chang, G.; Lee, C. C.; Huang, S. M.; Wang, B.; Bian, G.; Zheng, H.; Sanchez, D. S.; Belopolski, I.; Alidoust, N.; Neupane, M.; Bansil, A.; Jeng, H.-T.; Lin, H.; Zahid Hasan, M. Prediction of an Arc-Tunable Weyl Fermion Metallic State in $Mo_xW_{1-x}Te_2$. *Nat. Commun.* **2016,** *7*, 10639.

(8) Belopolski, I.; Xu, S. Y.; Ishida, Y.; Pan, X.; Yu, P.; Sanchez, D. S.; Zheng, H.; Neupane, M.; Alidoust, N.; Chang, G.; Chang, T. R.; Wu, Y.; Bian, G.; Huang, S. M.; Lee, C. C.; Mou, D.; Huang, L.; Song, Y.; Wang, B.; Wang, G.; *et al*. Fermi Arc Electronic Structure and Chern Numbers in the Type-II Weyl Semimetal Candidate $Mo_xW_{1-x}Te_2$. *Phys. Rev. B* **2016,** *94*, 085127.

(9) Belopolski, I.; Sanchez, D. S.; Ishida, Y.; Pan, X.; Yu, P.; Xu, S. Y.; Chang, G.; Chang, T.-R.; Zheng, H.; Alidoust, N.; Bian, G.; Neupane, M.; Huang, S. M.; Lee, C. C.; Song, Y.; Bu, H.; Wang, G.; Li, S.; Eda, G.; Jeng, H. T.; *et al*. Discovery of a New Type of Topological Weyl Fermion Semimetal State in $Mo_xW_{1-x}Te_2$. *Nat. Commun.* **2016,** *7*, 13643.

(10) Soluyanov, A. A.; Gresch, D.; Wang, Z.; Wu, Q.; Troyer, M.; Dai, X.; Bernevig, B. A. Type-II Weyl Semimetals. *Nature* **2015,** *527*, 495.

(11) Ali, M. N.; Xiong, J.; Flynn, S.; Tao, J.; Gibson, Q. D.; Schoop, L. M.; Liang, T.; Haldolaarachchige, N.; Hirschberger, M.; Ong, N. P.; Cava, R. J. Large, Non-Saturating Magnetoresistance in $WTe_2$. *Nature* **2014,** *514*, 205.

(12) Burkov, A. A.; Balents, L. Weyl Semimetal in a Topological Insulator Multilayer. *Phys. Rev. Lett.* **2011,** *107*, 127205.

(13) Huang, S. M.; Xu, S. Y.; Belopolski, I.; Lee, C. C.; Chang, G.; Wang, B.; Alidoust, N.; Bian, G.; Neupane, M.; Zhang, C.; Jia, S.; Bansil, A.; Lin, H.; Hasan, M. Z. A Weyl Fermion Semimetal with Surface Fermi Arcs in the Transition Metal Monopnictide TaAs Class. *Nat. Commun.* **2015,** *6*, 7373.

(14) Qi, Y.; Naumov, P. G.; Ali, M. N.; Rajamathi, C. R.; Schnelle, W.; Barkalov, O.; Hanfland, M.; Wu, S. C.; Shekhar, C.; Sun, Y.; Süß, V.; Schmidt, M.; Schwarz, U.; Pippel, E.; Werner, P.; Hillebrand, R.; Förster, T.; Kampert, E.; Parkin, S.; Cava, R. J.; *et al*. Superconductivity in Weyl Semimetal Candidate $MoTe_2$. *Nat. Commun.* **2016,** *7*, 11038.

(15) Tang, S.; Zhang, C.; Wong, D.; Pedramrazi, Z.; Tsai, H. Z.; Jia, C.; Moritz, B.; Claassen, M.; Ryu, H.; Kahn, S.; Jiang, J.; Yan, H.; Hashimoto, M.; Lu, D.; Moore, R. G.; Hwang, C.-C.; Hwang, C.; Hussain, Z.; Chen, Y.; Ugeda, M. M.; *et al*. Quantum Spin Hall State in Monolayer 1T'-$WTe_2$. *Nat. Phys.* **2017,** *13*, 683-687.

(16) Fatemi, V.; Wu, S.; Cao, Y.; Bretheau, L.; Gibson, Q. D.; Watanabe, K.; Taniguchi, T.; Cava, R. J.; Jarillo-Herrero, P. Electrically Tunable Low-Density Superconductivity in a Monolayer Topological Insulator. *Science* **2018,** *362*, 926-929.

(17) Lv, Y. Y.; Cao, L.; Li, X.; Zhang, B. B.; Wang, K.; Bin, P.; Ma, L.; Lin, D.; Yao, S. H.; Zhou, J.; Chen, Y. B.; Dong, S. T.; Liu, W.; Lu, M. H.; Chen, Y.; Chen, Y. F. Composition and Temperature-Dependent Phase Transition in Miscible Mo$_{1-x}$W$_x$Te$_2$ Single Crystals. *Sci. Rep-UK* **2017,** *7*, 44587.

(18) Sean, M. O.; Ryan, B.; Sergiy, K.; Irina, K.; Arunima, K. S.; Alina, B.; Francesca, T.; Jaydeep, J.; Iris, R. S.; Stephan, J. S.; Albert, V. D.; Patrick, M. V. The Structural Phases and Vibrational Properties of Mo$_{1-x}$W$_x$Te$_2$ Alloys. *2D Materials* **2017,** *4*, 045008.

(19) Aslan, O. B.; Datye, I. M.; Mleczko, M. J.; Sze Cheung, K.; Krylyuk, S.; Bruma, A.; Kalish, I.; Davydov, A. V.; Pop, E.; Heinz, T. F. Probing the Optical Properties and Strain-Tuning of Ultrathin Mo$_{1-x}$W$_x$Te$_2$. *Nano Lett.* **2018,** *18*, 2485-2491.

(20) Zhou, J.; Lin, J.; Huang, X.; Zhou, Y.; Chen, Y.; Xia, J.; Wang, H.; Xie, Y.; Yu, H.; Lei, J.; Wu, D.; Liu, F.; Fu, Q.; Zeng, Q.; Hsu, C. H.; Yang, C.; Lu, L.; Yu, T.; Shen, Z.; Lin, H.; *et al*. A Library of Atomically Thin Metal Chalcogenides. *Nature* **2018,** *556*, 355-359.

(21) Mathew, R. J.; Inbaraj, C. R. P.; Sankar, R.; Hudie, S. M.; Nikam, R. D.; Tseng, C. A.; Lee, C. H.; Chen, Y. T. High Unsaturated Room-Temperature Magnetoresistance in Phase-Engineered Mo$_x$W$_{1-x}$Te$_{2+\delta}$ Ultrathin Films. *J. Mater. Chem. C* **2019,** *7*, 10996-11004.

(22) Zhou, J.; Liu, F.; Lin, J.; Huang, X.; Xia, J.; Zhang, B.; Zeng, Q.; Wang, H.; Zhu, C.; Niu, L.; Wang, X.; Fu, W.; Yu, P.; Chang, T. R.; Hsu, C. H.; Wu, D.; Jeng, H. T.; Huang, Y.; Lin, H.; Shen, Z.; *et al*. Large-Area and High-Quality 2D Transition Metal Telluride. *Adv. Mater.* **2017,** *29*, 1603471.

(23) Jiang, Y. C.; Gao, J.; Wang, L. Raman Fingerprint for Semi-Metal WTe$_2$ Evolving from Bulk to Monolayer. *Sci. Rep-UK* **2016,** *6*, 19624.

(24) Rhodes, D.; Chenet, D. A.; Janicek, B. E.; Nyby, C.; Lin, Y.; Jin, W.; Edelberg, D.; Mannebach, E.; Finney, N.; Antony, A.; Schiros, T.; Klarr, T.; Mazzoni, A.; Chin, M.; Chiu, Y. c.; Zheng, W.; Zhang, Q. R.; Ernst, F.; Dadap, J. I.; Tong, X.; *et al*. Engineering the Structural and Electronic Phases of MoTe$_2$ through W Substitution. *Nano Lett.* **2017,** *17*, 1616-1622.

(25) Moulder, J. F; Stickle, W. F.; Sobol, P. E.; Bomben, K. D. *Handbook of X-Ray Photoelectron Spectroscopy*. Perkin-Elmer Corporation: Minnesota, USA, 1992; 40.

(26) Song, Q.; Wang, H.; Pan, X.; Xu, X.; Wang, Y.; Li, Y.; Song, F.; Wan, X.; Ye, Y.; Dai, L. Anomalous In-Plane Anisotropic Raman Response of Monoclinic Semimetal 1T′-MoTe$_2$. *Sci. Rep-UK* **2017,** *7*, 1758.

(27) Ma, X.; Guo, P.; Yi, C.; Yu, Q.; Zhang, A.; Ji, J.; Tian, Y.; Jin, F.; Wang, Y.; Liu, K.; Xia, T.; Shi, Y.; Zhang, Q. Raman Scattering in the Transition-Metal Dichalcogenides of 1T′-MoTe$_2$, T$_d$-MoTe$_2$, and T$_d$-WTe$_2$. *Phys. Rev. B* **2016,** *94*, 214105.

(28) Kim, Y.; Jhon, Y. I.; Park, J.; Kim, J. H.; Lee, S.; Jhon, Y. M. Anomalous Raman Scattering and Lattice Dynamics in Mono- and Few-Layer WTe$_2$. *Nanoscale* **2016,** *8*, 2309-2316.

(29) Zou, Y. C.; Chen, Z. G.; Liu, S.; Aso, K.; Zhang, C.; Kong, F.; Hong, M.; Matsumura, S.; Cho, K.; Zou, J. Atomic Insights into Phase Evolution in Ternary Transition-Metal Dichalcogenides Nanostructures. *Small* **2018,** *14*, 1800780.

(30) Chen, S. Y.; Goldstein, T.; Venkataraman, D.; Ramasubramaniam, A.; Yan, J. Activation of New Raman Modes by Inversion Symmetry Breaking in Type II Weyl Semimetal Candidate T′-MoTe$_2$. *Nano Lett.* **2016,** *16*, 5852-5860.

(31) Zandt, T.; Dwelk, H.; Janowitz, C.; Manzke, R. Quadratic Temperature Dependence up to 50 K of the Resistivity of Metallic MoTe$_2$. *J. Alloys Compd.* **2007,** *442*, 216-218.

(32) Tian, W.; Yu, W.; Liu, X.; Wang, Y.; Shi, J. A Review of the Characteristics, Synthesis, and Thermodynamics of Type-II Weyl Semimetal WTe$_2$. *Materials* **2018,** *11*, 1185.

(33) Wang, Y.; Liu, E.; Liu, H.; Pan, Y.; Zhang, L.; Zeng, J.; Fu, Y.; Wang, M.; Xu, K.; Huang, Z.; Wang, Z.; Lu, H. Z.; Xing, D.; Wang, B.; Wan, X.; Miao, F. Gate-Tunable Negative Longitudinal Magnetoresistance in the Predicted Type-II Weyl Semimetal WTe$_2$. *Nat. Commun.* **2016,** *7*, 13142.

(34) Tsen, A. W.; Hunt, B.; Kim, Y. D.; Yuan, Z. J.; Jia, S.; Cava, R. J.; Hone, J.; Kim, P.; Dean, C. R.; Pasupathy, A. N. Nature of the Quantum Metal in a Two-Dimensional Crystalline Superconductor. *Nat. Phys.* **2016,** *12*, 208-212.

(35) Xu, C.; Wang, L.; Liu, Z.; Chen, L.; Guo, J.; Kang, N.; Ma, X. L.; Cheng, H. M.; Ren, W. Large-Area High-Quality 2D Ultrathin Mo$_2$C Superconducting Crystals. *Nat. Mater.* **2015,** *14*, 1135-1141.

(36) Cui, J.; Li, P.; Zhou, J.; He, W. Y.; Huang, X.; Yi, J.; Fan, J.; Ji, Z.; Jing, X.; Qu, F.; Cheng, Z. G.; Yang, C.; Lu, L.; Suenaga, K.; Liu, J.; Law, K. T.; Lin, J.; Liu, Z.; Liu, G. Transport Evidence of Asymmetric Spin–Orbit Coupling in Few-Layer Superconducting 1T$_d$-MoTe$_2$. *Nat. Commun.* **2019,** *10*, 2044.

(37) Tinkham, M.. *Introduction to Superconductivity*, 2nd; Courier Corporation: New York, USA, 2004.

(38) Yang, C.; Liu, Y.; Wang, Y.; Feng, L.; He, Q.; Sun, J.; Tang, Y.; Wu, C.; Xiong, J.; Zhang, W.; Lin, X.; Yao, H.; Liu, H.; Fernandes, G.; Xu, J.; Valles, J. M.; Wang, J.; Li, Y. Intermediate Bosonic Metallic State in the Superconductor-Insulator Transition. *Science* **2019**, *366*, 1505-1509.

(39) Zhang, C.; Fan, Y.; Chen, Q.; Wang, T.; Liu, X.; Li, Q.; Yin, Y.; Li, X. Quantum Griffiths Singularities in TiO Superconducting Thin Films with Insulating Normal States. *NPG Asia Mater.* **2019,** *11*, 76.

(40) Saito, Y.; Nojima, T.; Iwasa, Y. Quantum Phase Transitions in Highly Crystalline Two-Dimensional Superconductors. *Nat. Commun.* **2018,** *9*, 778.

(41) Zheliuk, O.; Lu, J. M.; Chen, Q. H.; Yumin, A. A. E.; Golightly, S.; Ye, J. T. Josephson Coupled Ising Pairing Induced in Suspended $MoS_2$ Bilayers by Double-Side Ionic Gating. *Nat. Nanotechnol.* **2019,** *14*, 1123-1128.

(42) Zeng, J.; Liu, E.; Fu, Y.; Chen, Z.; Pan, C.; Wang, C.; Wang, M.; Wang, Y.; Xu, K.; Cai, S.; Yan, X.; Wang, Y.; Liu, X.; Wang, P.; Liang, S. J.; Cui, Y.; Hwang, H. Y.; Yuan, H.; Miao, F. Gate-Induced Interfacial Superconductivity in $1T-SnSe_2$. *Nano Lett.* **2018,** *18*, 1410-1415.

(43) Huang, X.; Zhao, L.; Long, Y.; Wang, P.; Chen, D.; Yang, Z.; Liang, H.; Xue, M.; Weng, H.; Fang, Z.; Dai, X.; Chen, G. Observation of the Chiral-Anomaly-Induced Negative Magnetoresistance in 3D Weyl Semimetal TaAs. *Phys. Rev. X* **2015,** *5*, 031023.

(44) Son, D. T.; Spivak, B. Z. Chiral Anomaly and Classical Negative Magnetoresistance of Weyl Metals. *Phys. Rev. B* **2013,** *88*, 104412.

(45) Zhang, C. L.; Xu, S. Y.; Belopolski, I.; Yuan, Z.; Lin, Z.; Tong, B.; Bian, G.; Alidoust, N.; Lee, C. C.; Huang, S. M.; Chang, T. R.; Chang, G.; Hsu, C. H.; Jeng, H. T.; Neupane, M.; Sanchez, D. S.; Zheng, H.; Wang, J.; Lin, H.; Zhang, C.; *et al*. Signatures of the Adler–Bell–Jackiw Chiral Anomaly in a Weyl Fermion Semimetal. *Nat. Commun.* **2016,** *7*, 10735.

(46) Li, P.; Wen, Y.; He, X.; Zhang, Q.; Xia, C.; Yu, Z. M.; Yang, S. A.; Zhu, Z.; Alshareef, H. N.; Zhang, X. X. Evidence for Topological Type-II Weyl Semimetal $WTe_2$. *Nat. Commun.* **2017,** *8*, 2150.

(47) Hikami, S.; Larkin, A. I.; Nagaoka, Y. Spin-Orbit Interaction and Magnetoresistance in the Two Dimensional Random System. *Prog. Theor. Phys.* **1980,** *63*, 707-710.

(48) Wang, Q.; Yu, P.; Huang, X.; Fan, J.; Jing, X.; Ji, Z.; Liu, Z.; Liu, G.; Yang, C.; Lu, L. Observation of Weak Anti-Localization and Electron-Electron Interaction on Few-Layer 1T′-$MoTe_2$ Thin Films. *Chin. Phys. Lett.* **2018,** *35*, 077303.